\documentclass[showpacs,amsmath,amssymb,twocolumn]{revtex4}

\newcommand{\be}{\begin{equation}}
\newcommand{\ee}{\end{equation}}
\newcommand{\ba}{\begin{array}}
\newcommand{\ea}{\end{array}}
\newcommand{\bea}{\begin{eqnarray}}
\newcommand{\eea}{\end{eqnarray}}
\newcommand{\bma}{\begin{matrix}}
\newcommand{\ema}{\end{matrix}}
\newcommand{\bpm}{\begin{pmatrix}}
\newcommand{\epm}{\end{pmatrix}}
\newcommand{\nn}{\nonumber}

\newcommand{\ov}{\overline}

\begin{document}

\title{SuperHiggs Mechanism in String Theory}

\author{Jonathan Bagger}
\email{bagger@jhu.edu}
\affiliation{Department of Physics and Astronomy, Johns Hopkins University, 3400 North Charles Street, Baltimore, MD 21218, USA}
\author{Ioannis Giannakis}
\email{giannak@theory.rockefeller.edu}
\affiliation{Department of Physics, Rockefeller University, 1230 York Avenue, New York, NY 10021, USA}

\date{\today}

\begin{abstract}
We exhibit the superHiggs effect in heterotic string theory by turning on a background NS-NS field and deforming the BRST operator consistent with superconformal invariance.  The NS-NS field spontaneously breaks spacetime supersymmetry.  We show how the gravitini and the physical dilatini gain mass by eating the would-be Goldstone fermions.
\end{abstract}

\maketitle


\section{Introduction}

String theory is the most ambitious and most promising attempt to incorporate gravity into quantum mechanics.  The theory possesses a large number of symmetries, including supersymmetry, gauge symmetry and coordinate invariance.  At low energies, many of the symmetries are broken by the vacuum.  At ultra-high energies, beyond the Planck scale, the results of \cite{gross} suggest that the full symmetry group is infinite-dimensional.

How can we investigate the symmetry structure of the theory?  Evans and Ovrut developed an elegant approach in which string symmetries are generated by inner automorphisms of the superconformal operator algebra \cite{eo}.  The method treats unbroken and spontaneously broken symmetries on exactly the same footing.  In recent work, we used this formalism to illustrate the Higgs mechanism in string theory \cite{ioannis}.

In this paper we extend these results to the case of spontaneously broken supersymmetry.  In sections 2 and 3 we focus on string propagation in flat Minkowski space, in the presence of a non-trivial but infinitesimal NS-NS two-form $B$.  We assume the $B$-field background satisfies the string theory equations of motion, and derive the string-theory equations of motion for the gravitino and the dilatino fields.  We also find the spacetime supersymmetry generator in this background.  We then use the supersymmetry generator to derive the spacetime supersymmetry transformations of the gravitino and the dilatino fields.

In section 4 we use these results to illustrate the superHiggs mechanism in string theory.  We study a simple model in which spacetime is compactified on $M^7 \times T^3$, with a constant $H=dB$ flux in the compact dimensions.  For zero flux, the seven-dimensional theory has $N=2$ spacetime supersymmetry, with two massless gravitini and eight massless dilatini.  The non-zero $H$ spontaneously breaks the supersymmetry. The gravitini and the dilatini obey coupled
equations of motion.
We show that the would-be Goldstone fermions can be eliminated by a supersymmetry transformation, and that in the unitary gauge, the gravitini and the remaining six dilatini obey massive equations of motion.
Aspects of supersymmetry breaking in string theory were
discussed previously in \cite{antoniadis}. 

\section{Nilpotent Deformations and Equations of Motion}

\subsection{Heterotic String in Minkowski Space}

To fix notation, we first describe the heterotic string in flat Minkowski space.  We start with the left- and right-moving BRST operators $Q$ and ${\overline Q}$, which are given by
\bea
Q&=&{\int}d{\sigma} \left(  c \left( T+{\partial}cb\right)
-\frac{1}{2}{\gamma}T_{F} -\frac{1}{4}b{\gamma}^2 \right) \nn \\
{\overline Q}&=&{\int}d{\sigma}\ {\overline c} \left({\overline T}
+{\ov\partial}{\ov c}{\ov b} \right),
\label{eqvios}
\eea
where the world-sheet stress-energy tensor is
\bea
T&=&\frac{1}{2} \eta_{\mu\nu} {\partial}X^\mu{\partial}X^\nu
+\frac{1}{2}\eta_{\mu\nu}{\psi^\mu}{\partial\psi}^\nu
-\frac{3}{2}{\partial\beta}{\gamma}-\frac{1}{2}{\beta}{\partial\gamma}\nn\\
{\overline T}&=&\frac{1}{2} \eta_{\mu\nu}
{\overline\partial}X^\mu{\overline\partial}X^\nu,
\eea
the world-sheet supercurrent is
\be
T_F = \frac{1}{2} \eta_{\mu\nu} {\psi^\mu}{\partial} X^\nu,
\ee
and $b,c$ and $\beta,\gamma$ are the world-sheet ghosts (together with their conjugates).  The bosonic ghosts $\beta$ and $\gamma$ can be bosonized as follows,
\be
\beta=e^{-\phi}{\partial\xi}, \quad  \gamma=e^{\phi}{\eta},
\ee
where $\xi$ and $\eta$ are conjugate fermions of dimension 0 and 1, and $\phi$ is a chiral boson.

An infinitesimal gravitino excitation deforms the BRST operators as follows,
\be
Q \rightarrow Q + \delta Q, \quad {\overline Q}
\rightarrow {\overline Q}+\delta {\overline Q},
\label{eqkrokos}
\ee
where
\bea
\delta Q &=& {\int}d{\sigma}\ c \left( {\Psi}^{\alpha}_{\mu}S_{\alpha}e^{-{\frac{\phi}{2}}}
{\ov\partial}X^{\mu} \right)\nn\\[2mm]
\delta \ov{Q} &=& {\int}d{\sigma}\ \ov{c} \left( {\Psi}^{\alpha}_{\mu}S_{\alpha}e^{-{\frac{\phi}{2}}}
{\ov\partial}X^{\mu} \right),
\label{eqcreos}
\eea
and $S_\alpha$ is a world-sheet spin field.  Nilpotency requires
\be
\lbrace Q, \delta Q \rbrace =   \lbrace
{\overline Q},
\delta {\overline Q} \rbrace=   \lbrace Q,
\delta {\overline Q}
\rbrace+\lbrace {\overline Q}, \delta Q \rbrace=0,
\label{eqkwmh}
\ee
which in turn imposes the following restrictions on the field $\Psi^{\alpha}_{\mu}$,
\be
\quad ({\gamma}^{\mu})_{\alpha\beta}{\partial_{\mu}}
{\Psi^{\beta}_{\nu}}=0, \quad {\partial^{\mu}}
\Psi^{\alpha}_{\mu}=0.
\label{eqasev}
\ee

The relations (\ref{eqasev}) contain an equation of motion and a gauge condition.  However, the gauge condition does not separate the physical degrees of freedom.  As written, $\Psi^{\alpha}_{\mu}$ contains a spin-$\frac{3}{2}$ gravitino and a spin-$\frac{1}{2}$ dilatino.  To separate the fields, we write ${\Psi}_{\mu}^{\alpha}= {\chi}_{\mu}^{\alpha}+({\gamma_{\mu}})^{\alpha\beta}{\lambda}_{\beta}$, where $(\gamma^{\mu})_{\alpha\beta}{\chi}_{\mu}^{\beta}=0$.  Equations (\ref{eqasev}) then become
\bea
({\gamma}^{\mu})_{\alpha\beta}{\partial_{\mu}}
{\chi}_{\nu}^{\beta} =2{\partial_{\nu}}{\lambda_\alpha},
\quad (\gamma^{\mu})_{\alpha\beta}{\chi}_{\mu}^{\beta}=0,\nn\\
\quad {\partial^{\mu}}
{\chi}_{\mu}^{\alpha} =0, \quad
({\gamma}^{\mu})_{\alpha\beta}{\partial_{\mu}}
{\lambda}^{\beta} =0.
\label{eqasxiuf}
\eea
These are coupled equations of motion for a massless spin-$\frac{3}{ 2}$ gravitino and a massless spin-$\frac{1}{ 2}$ dilatino, in a generalized Lorentz gauge.

\subsection{Heterotic String in a NS-NS Background}

To study supersymmetry breaking, we will work in a background with an infinitesimal NS-NS gauge field $B_{\mu\nu}$ \cite{bagger}.  This deforms the BRST 
operators as follows,
\be
Q \rightarrow Q + \delta Q', \quad {\overline Q}
\rightarrow {\overline Q}+\delta {\overline Q}',
\ee
where
\bea
\delta Q' &=&{\int}d{\sigma}\Big( c \left( B_{\mu\nu}
{\partial}X^{\mu}{\ov\partial}X^{\nu}
+ {\partial_\kappa}
B_{\mu\nu}{\psi}^{\kappa}{\psi}^{\mu}{\ov\partial}X^{\nu}\right)\nn\\
&&
-\frac{1}{ 2}{\gamma}B_{\mu\nu}{\psi}^{\mu}{\ov\partial}X^{\nu}\Big) \nn\\
\delta{\ov Q}' &=&{\int}d{\sigma}\ {\ov c} \left(B_{\mu\nu}
{\partial}X^{\mu}{\ov\partial}X^{\nu}+{\partial_\kappa}
B_{\mu\nu}{\psi}^{\kappa}{\psi}^{\mu}{\ov\partial}X^{\nu}\right).\nn\\
\label{eqcreos2}
\eea
Nilpotency requires
\be
\lbrace Q, \delta Q' \rbrace =  \lbrace
{\overline Q},
\delta {\overline Q}' \rbrace=  \lbrace Q,
\delta {\overline Q}'
\rbrace+\lbrace {\overline Q}, \  Q' \rbrace=0,
\label{eqkwmh2}
\ee
which imposes an equation of motion and a gauge condition for the $B_{\mu\nu}$ field,
\be
\Box B_{\mu\nu}=0, \quad \partial^\mu  B_{\mu\nu}=0.
\ee

As in flat space, a gravitino excitation also deforms the BRST operators,
\be
Q \rightarrow Q + \delta Q' + \delta Q'' \ , \quad {\overline Q}
\rightarrow {\overline Q}+\delta {\overline Q}' + \delta {\overline Q}'',
\label{eqkrokos3}
\ee
where
\begin{widetext}
\bea
\delta Q'' & =& {\int}d{\sigma}\ c\ {\Psi}^{\alpha}_{\mu}
\left( S_{\alpha}e^{-{\frac{\phi}{2}}} {\ov\partial}X^{\mu}
+\frac{1}{4} ({\gamma}^{\kappa\nu})_{\alpha}^{\beta}B_{\kappa\nu}
S_{\beta}e^{-{\frac{\phi}{ 2}}}{\ov\partial}X^{\mu}
-\frac{1}{ 2}
B^{\mu\nu}S_{\alpha}e^{-{\frac{\phi}{ 2}}}({\partial}X_{\nu}
-{\ov\partial}X_{\nu}) -\frac{1}{ 2}{\partial^\kappa}B^{\mu\nu}
S_{\alpha}e^{-{\frac{\phi}{ 2}}}{\psi_\kappa}{\psi_\nu}  \right) \nn\\[2mm]
\delta {\ov Q}'' & =& {\int}d{\sigma}\ \ov{c}\ {\Psi}^{\alpha}_{\mu}
\left( S_{\alpha}e^{-{\frac{\phi}{2}}} {\ov\partial}X^{\mu}
+\frac{1}{4} ({\gamma}^{\kappa\nu})_{\alpha}^{\beta}B_{\kappa\nu}
S_{\beta}e^{-{\frac{\phi}{ 2}}}{\ov\partial}X^{\mu} 
-\frac{1}{ 2}
B^{\mu\nu}S_{\alpha}e^{-{\frac{\phi}{ 2}}}({\partial}X_{\nu}
-{\ov\partial}X_{\nu}) -\frac{1}{ 2}{\partial^\kappa}B^{\mu\nu}
S_{\alpha}e^{-{\frac{\phi}{ 2}}}{\psi_\kappa}{\psi_\nu}  \right) ,\nn\\
\label{eqrewo}
\eea
and we work to first order in $B_{\mu\nu}$ and $\Psi^\alpha_\mu$.  Nilpotency imposes an additional equation of motion
\be
({\gamma}^\mu)_{\alpha\beta}{\partial_{\mu}}
{\Psi^{\beta}_{\nu}}-({\gamma}^\mu)_{\alpha\beta}
H_{\mu\nu}^{\kappa}\Psi^{\beta}_{\kappa}
+\frac{1}{ 6}({\gamma}^{\mu\kappa\rho})_{\alpha\beta}
H_{\mu\kappa\rho}{\Psi^{\beta}_{\nu}}=0,
\label{eqriddle}
\ee
and gauge condition,
\be
{\partial^\mu}{\Psi^{\alpha}_{\mu}}
+\frac{1}{ 2}({\gamma}_{\kappa\nu})_{\beta}^{\alpha}
H^{\mu\kappa\nu}{\Psi^{\beta}_{\mu}}=0,
\label{eqfiddle}
\ee
where $H_{\mu\kappa\nu}=\frac{1}{ 2}({\partial_\nu}B_{\mu\kappa} +{\partial_\mu}B_{\kappa\nu}+{\partial_\kappa}B_{\nu\mu})$.  Writing ${\Psi^{\alpha}_{\mu}}$ in terms of its trace and traceless part, and substituting into (\ref{eqriddle}) and (\ref{eqfiddle}), we find equations of motion for the gravitino $\chi^{\alpha}_\mu$,
\bea
({\gamma}^\mu)_{\alpha\beta}{\partial_{\mu}}
{\chi_{\nu}^{\beta}}-({\gamma}^\mu)_{\alpha\beta}
H_{\mu\nu}{}^{\kappa}{\chi_{\kappa}^{\beta}}
+\frac{1}{ 6}({\gamma}^{\mu\kappa\rho})_{\alpha\beta}
H_{\mu\kappa\rho}{\chi_{\nu}^{\beta}}+\frac{2}{ D}
({\gamma}_{\nu}^{\kappa\mu})_{\alpha\beta}H_{\kappa\mu}{}^{\rho}
{\chi_{\rho}^{\beta}}
+\frac{4}{ D}({\gamma}^\mu)_{\alpha\beta}
H_{\mu\nu}{}^\rho{\chi_{\rho}^\beta}\nn\\
=
{\partial_\nu}{\lambda}_{\alpha}-\frac{3}{ 2}
({\gamma}^{\kappa\mu})_{\alpha}^{\beta}H_{\kappa\mu\nu}{\lambda}_\beta
-\frac{2}{ D}({\gamma}_{\nu}{\gamma^{\mu\kappa\rho}})_{\alpha}^{\beta}
H_{\mu\kappa\rho}{\lambda}_{\beta},
\label{eqfiddler1}
\eea
the dilatino $\lambda_\alpha$,
\be
({\gamma}^\mu)^{\alpha\beta}{\partial_{\mu}}
{\lambda_{\beta}}
+\frac{1}{ 6}({\gamma}^{\mu\kappa\rho})^{\alpha\beta}
H_{\mu\kappa\rho}{\lambda_{\beta}}=-\frac{2}{ D}
({\gamma}^{\kappa\rho})^{\alpha}_{\beta}H_{\kappa\rho}{}^{\mu}
{\chi_{\mu}^{\beta}}+\frac{2}{ D}
({\gamma}^{\kappa\rho}{\gamma_\mu})^{\alpha\beta}H_{\kappa\rho}{}^{\mu}
{\lambda_{\beta}},
\label{eqfiddler2}
\ee
together with the gauge condition
\be
{\partial^\mu}{\chi_{\mu}^{\alpha}}=\frac{4+D}{ 2D}
({\gamma}^{\kappa\rho})^{\alpha}_{\beta}H_{\kappa\rho}^{\mu}
{\chi_{\mu}^{\beta}}+\frac{6+D}{ 3D}
({\gamma}^{\mu\kappa\rho})^{\alpha\beta}H_{\mu\kappa\rho}
{\lambda_{\beta}},
\label{eqfiddler3}
\ee
\end{widetext}
where $D$ is the dimension of spacetime.  These are the generalizations of (\ref{eqasxiuf}) in the $B_{\mu\nu}$ background.

\section{Spacetime Supersymmetry}

In string theory, spacetime symmetries correspond to inner automorphisms of the operator algebra.  They are generated by infinitesimal operators $h$,
\be
i[h, {\cal O}]=\delta {\cal O},
\label{eqrios}
\ee
where ${\cal O}$ is any operator in the theory.  When ${\cal O}$ is $Q$ or $\ov Q$, the deformed BRST charges $\delta Q$ and $\delta \ov Q$ automatically satisfy the deformation equations (\ref{eqkwmh}) because of the Bianchi identity.

Following Evans and Ovrut \cite{eo}, we define a {\it canonical deformation} to be generated by an infinitesimal operator $h$ that is the sum of zero modes of $(1, 0)$ and $(0, 1)$ primary fields.  Such a deformation preserves the gauge of the spacetime fields.  For the heterotic string, the operator that generates a supersymmetry transformation about flat spacetime is \cite{friedan}
\be
h={\int}d{\sigma}{\epsilon}^{\alpha}S_{\alpha}e^{-{\frac{\phi}{ 2}}}.
\label{eqperez}
\ee
The integrand is of dimension $(1, 0)$ provided the transformation parameter $\epsilon^{\alpha}$ satisfies
\be
({\gamma}^{\mu})_{\alpha\beta}{\partial_{\mu}}
{\epsilon^{\beta}}=0.
\label{eqaseme}
\ee
In this case, $h$ generates a canonical deformation.

The supersymmetry transformation of the gravitino can be found by commuting $h$ with the BRST operator $Q$.  For the case at hand, we find
\be
i[h, Q]= \delta Q = \int d\sigma \ c \left( {\partial_\mu}{\epsilon}^{\alpha}
S_{\alpha}e^{-{\frac{\phi}{2}}}{\ov\partial}X^{\mu}\right) ,
\label{eqviacom}
\ee
and likewise for $ \delta \ov Q$.
Comparing with (\ref{eqcreos}), we see that the commutator (\ref{eqviacom}) describes a deformation of the ${\Psi}^{\alpha}_{\lambda}$ field.  In this way $Q$ and $\ov Q$ generate a flat-space supersymmetry transformation, ${\delta}{\Psi}^{\alpha}_{\lambda}={\partial_\lambda} {\epsilon}^{\alpha}$.

The supersymmetry generator $h$ deforms in the  $B_{\mu\nu}$ background,
\be
h \rightarrow h + \delta h ={\int}d{\sigma}\ {\epsilon}^{\alpha}\left(
S_{\alpha}e^{-{\frac{\phi}{2}}}+\frac{1}{4}
({\gamma}^{\mu\nu})_{\alpha}^{\beta}B_{\mu\nu}
S_{\beta}e^{-{\frac{\phi}{ 2}}}\right).
\label{eqcit}
\ee
The corresponding deformation is canonical if $\epsilon^{\alpha}$ satisfies the following constraint,
\be
({\gamma}^{\mu})_{\alpha\beta}{\partial_\mu}{\epsilon}^{\alpha}
+\frac{1}{ 6}({\gamma}^{\mu\nu\kappa})_{\alpha\beta}
H_{\mu\nu\kappa}{\epsilon}^{\alpha}=0.
\label{eqdirac}
\ee
This is the Dirac equation for $\epsilon^\alpha$ in the $B_{\mu\nu}$ background.

To find the gravitino transformation in this background, we compute the commutator of $h + \delta h$ with $Q+\delta Q' $.  This gives
\begin{widetext}
\bea
i[h+ \delta h, Q+\delta Q']&=& \int d\sigma \ c \left({\partial_{\mu}}{\epsilon}^{\alpha}
+\frac{1}{ 2}({\gamma^{\rho\lambda}})_{\beta}^{\alpha}
H_{\mu\rho\lambda}{\epsilon}^{\beta}\right)
\left( S_{\alpha}e^{-{\frac{\phi}{2}}} {\ov\partial}X^{\mu}\right.\nn\\
&&+\frac{1}{4} ({\gamma}^{\nu\kappa})_{\alpha}^{\beta}B_{\nu\kappa}
S_{\beta}e^{-{\frac{\phi}{ 2}}}{\ov\partial}X^{\mu} -\frac{1}{ 2}
B^{\mu\kappa}S_{\alpha}e^{-{\frac{\phi}{ 2}}}({\partial}X_{\kappa}-{\ov\partial}X_{\kappa})\nn\\
&& \left.
 -\frac{1}{ 2}{\partial^\rho}B^{\mu\kappa}
S_{\alpha}e^{-{\frac{\phi}{ 2}}}{\psi_\rho}{\psi_\kappa}  \right).
\label{eqrewod}
\eea
Comparing with (\ref{eqrewo}), we can read off the ${\Psi}^{\alpha}_{\lambda}$ transformation in the $B_{\mu\nu}$  background,
\be
{\delta}{\Psi}_{\mu}^{\alpha}={\partial_{\mu}}{\epsilon}^{\alpha}
+\frac{1}{ 2}({\gamma^{\nu\kappa}})_{\beta}^{\alpha}
H_{\mu\nu\kappa}{\epsilon}^{\beta}.
\label{eqswift}
\ee
Decomposing ${\Psi}_{\mu}^{\alpha}=\chi_\mu^{\alpha}+({\gamma_\mu})^{\alpha\beta}{\lambda}_\beta$, we find the transformation properties of the gravitino and the dilatino,
\bea
{\delta}\chi_\mu^{\alpha}&=&{\partial_\mu}{\epsilon^\alpha}
+\frac{1}{ 2}({\gamma^{\nu\kappa}})_{\beta}^{\alpha}H_{\mu\nu\kappa}
{\epsilon^\beta}-\frac{2}{ 3D}({\gamma}_{\mu}
{\gamma^{\nu\kappa\rho}})^{\alpha}_{\beta}H_{\nu\kappa\rho}
{\epsilon}^{\beta}\nn\\
{\delta}{\lambda}_{\alpha}&=&\frac{2}{ 3D}
({\gamma^{\mu\nu\kappa}})_{\alpha\beta}H_{\mu\nu\kappa}
{\epsilon}^{\beta}.
\label{eqgolding}
\eea
\end{widetext}
There are precisely the transformations of ten-dimensional supergravity, derived directly from string theory.

\section{The SuperHiggs Mechanism}

In the previous sections we studied string theory in the presence of a $B_{\mu\nu}$ field that pervades all of spacetime. In this section we focus on string propagation on $M^7 \times T^3$, where the $B_{\mu\nu}$ field is restricted to $T^3$.  We will see that the $B_{\mu\nu}$ field spontaneously breaks the supersymmetry on $M^7$.

We start by fixing the notation.  We take the spacetime coordinates to be $\{ X^\mu, X^i\}$, where $\mu=0, \cdots 6$; and $i=7, 8, 9$.  We decompose the ten-dimensional gamma matrices in direct product fashion,
\be
\Gamma^{\mu}=\frac{i} {\sqrt 2}
({\gamma}^{\mu}\otimes 1 \otimes {\sigma^1}), \quad
\Gamma^{i}=\frac{i} {\sqrt 2}(
1\otimes {\sigma^i} \otimes {\sigma^2}),
\label{eqvitiro}
\ee
where the $\gamma^\mu$ satisfy the Clifford algebra in $7$ dimensions and the $\sigma^i$ are ordinary Pauli matrices.  With these conventions, the ten-dimensional gravitino splits into two seven-dimensional gravitini and two seven-dimensional dilatini, ${\Psi}^{{\alpha}a}_{\mu} = {\chi}_{\mu}^{\alpha a} +({\gamma_{\mu}})^{\alpha\beta}{\lambda}^a_{\beta}$,
together with six additional seven-dimensional dilatini, $\Psi^{\alpha a}_{i}= -i \lambda^{\alpha a}_{i}$, where $\alpha=1, \cdots 8$ and $a=1,2$.

Let us focus on the background in which $H_{ijk}=2 m \epsilon_{ijk}$.  Nilpotency of the BRST operators implies that the dilatini ${\lambda}^{\alpha a}_{i}$ obey the following equations of motion,
\be
(\gamma^\mu)_{\alpha\beta}  \partial_\mu
{\lambda}^{\beta a}_{i}+2im {\epsilon_{ij}{}^k} ({\sigma^j})^{ab}
\lambda^{b}_{\alpha k}
+m {\lambda^{a}_{\alpha i}}=0.
\label{eqnew}
\ee
It also imposes equations of motion
\be
(\gamma^\mu)_{\alpha\beta}{\partial_\mu}{\Psi}^{\beta a}_{\lambda}
+m {\Psi}^{a}_{\alpha \lambda}=0,
\label{eqdefo}
\ee
and gauge conditions
\be
{\partial^\mu}{\Psi^{{\alpha}a}_\mu}
-im({\sigma^i})^{ab} {\Psi^{{\alpha}b}_{i}}=0.
\label{eqretina}
\ee
on the ${\Psi}^{{\alpha}a}_{\lambda}$.  In terms of gravitini and dilatini parts, eqs.~(\ref{eqdefo}) and (\ref{eqretina}) can be written as follows,
\bea
(\gamma^\mu)_{\alpha\beta}{\partial_\mu}{\chi}^{{\beta}a}_{\nu}
+m
{\chi}^{a}_{\alpha\nu}&=&-2{\partial_\nu}{\lambda^{a}_{\beta}}
+\frac{2m}{ 7}(\gamma_\nu)_{\alpha\beta}(\sigma^i)^{ab}
{\lambda^{{\beta}b}_{i}}\nn\\
(\gamma^\mu)_{\alpha\beta}{\partial_\mu}{\lambda}^{\beta a}
-m
{\lambda}^{a}_{\alpha}&=&\frac{2m}{ 7}(\sigma^i)^{ab}
{\lambda^{b}_{\alpha i}}\nn\\
{\partial^\mu}{\chi^{{\alpha}a}_{\mu}}+
m
{\lambda}^{\alpha a}&=&\frac{5m} {7}(\sigma^i)^{ab}
{\lambda^{{\alpha}b}_{i}} .
\label{eqdeforma}
\eea

The gravitini and dilatini obey coupled equations of motion.  Multiplying eq.\ (\ref{eqnew}) by $\sigma^i$ and using eqs.\ (\ref{eqdeforma}), we find
\begin{widetext}
\bea
({\gamma^\mu})_{\alpha\beta} {\partial_\mu}(\chi_\nu^{\beta a}+
\frac{1}{ 7}({\gamma_\nu})^{\beta\gamma}
({\sigma^i})^{ab}
{\lambda_{\gamma i}^b})
+m(\chi_\nu^{\alpha a}+
\frac{1}{ 7}({\gamma_\nu})^{\alpha\beta}(\sigma^i)^{ab}
\lambda_{\beta i}^b)&=&-2{\partial_\nu}(\lambda^a_\alpha-\frac{1}{ 7}({\sigma^i})^{ab}
\lambda^b_{\alpha i}) \nn\\
({\gamma^\mu})_{\alpha\beta} {\partial_\mu}(\lambda^{\beta a}-\frac{1}{ 7}
(\sigma^i)^{ab}
{\lambda^{\beta b}_{ i}})-m(\lambda^a_\alpha-\frac{1}{ 7}({\sigma^i})^{ab}
{\lambda^b_{\alpha i}})&=&0 \nn\\
{\partial^\mu}({\chi^{\alpha a}_\mu}+\frac{1}{ 7}{(\gamma_\mu)^{\alpha\beta}}
({\sigma^i})^{ab}\lambda^b_{\beta i})+m(\lambda^{\alpha a}-\frac{1}{ 7}
({\sigma^i})^{ab}{\lambda^{\alpha b}_i})-m
({\sigma^i})^{ab}{\lambda^{\alpha b}_i} &=& 0.
\label{eqvirio}
\eea
\end{widetext}
The form of these equations suggests the following change of variables,
\bea
\chi_\mu^{\prime\, \alpha a} &=& \chi_\mu^{\alpha a}+
\frac{1}{ 7}({\gamma_\mu})^{\alpha\beta}
({\sigma^i})^{ab}
{\lambda_{\beta i}^b} \nn\\
\lambda_{i}^{\prime\, \alpha a} &=& \lambda_{i}^{\alpha a}
-\frac{1}{3}({\sigma_i})^{ab}
({\sigma^j})^{bc}\lambda_{j}^{\alpha c} \nn\\
\lambda^{\prime\, a}_{1\alpha} &=& \lambda^a_\alpha
-\frac{1}{ 7}({\sigma^i})^{ab}
\lambda^b_{\alpha i} \nn\\
\lambda^{\prime\, a}_{2\alpha} &=& \lambda^a_\alpha
-\frac{1}{21}({\sigma^i})^{ab}
\lambda^b_{\alpha i}.
\label{redefs}
\eea
In terms of the primed variables, the equations and gauge conditions become
\bea
({\gamma^\mu})_{\alpha\beta} {\partial_\mu}\chi_\nu^{\prime\, \beta a}
+m\chi_\nu^{\prime\, \alpha a}
&=&-2{\partial_\nu}\lambda^{\prime\, a}_{1\alpha} \nn\\[1mm]
{\partial^\mu}{\chi^{\prime\,\alpha a}_\mu}
+m\lambda_{1}^{\prime\,\alpha a}-m
({\gamma^\mu})^{\alpha\beta}{\chi^{\prime\, a}_{\beta\mu}} &=& 0 \nn\\[1mm]
(\gamma^\mu)_{\alpha\beta}  \partial_\mu
{\lambda}^{\prime\, \beta a}_{i}
-3m {\lambda^{\prime\, a}_{\alpha i}}&=&0 \nn\\[1mm]
({\gamma^\mu})_{\alpha\beta} {\partial_\mu}\lambda_{1}^{\prime\,\beta a}
-m\lambda^{\prime\, a}_{1\alpha} &=& 0 \nn\\[1mm]
({\gamma^\mu})_{\alpha\beta} {\partial_\mu}\lambda_{2}^{\prime\,\beta a}
-3m\lambda^{\prime\, a}_{2\alpha}+2m\lambda^{\prime\, a}_{1\alpha}&=&0.
\label{eqvirio2}
\eea

To properly interpret these relations, we need to find the supersymmetry transformations in this background.  The supersymmetry generator is
\be
h={\int}d{\sigma}{\epsilon}^{{\alpha}a}\left(
S^a_{{\alpha}}e^{-{\frac{\phi}{ 2}}}+\frac{1}{
4}({\sigma}^{ij})^{ab}B_{ij}
S^b_{{\alpha}}e^{-{\frac{\phi}{ 2}}}\right),
\label{eqcite}
\ee
where the transformation parameters ${\epsilon}^{{\alpha}a}$ obey the following conditions,
\be
({\gamma}^{\mu})_{\alpha\beta}{\partial_\mu}{\epsilon}^{{\beta}a}
+m
{\epsilon}_\alpha^a =0, \quad
\Box {\epsilon}_\alpha^a=0.
\label{eqnewton}
\ee
Following the arguments of section 3, we can derive the supersymmetry transformations of the gravitini and dilatini,
\bea
{\delta}{\chi^{{\alpha}a}_{\mu}}&=&{\partial_\mu}
{\epsilon^{{\alpha}a}}
-\frac{m}{7}({\gamma_\mu})^{\alpha\beta}
{\epsilon}^{a}_{\beta}\nn\\
{\delta}{\lambda^{\alpha a}}&=&\frac{m}{7}
{\epsilon}^{{\alpha}a}\nn\\
{\delta}{\lambda_i^{{\alpha}a}}&=&m(\sigma_i)^{ab}
{\epsilon^{{\alpha}b}} .
\label{eqfinal}
\eea
The dilatini transform non-linearly under supersymmetry transformations.

We now have what we need to interpret eqs.~(\ref{eqvirio2}).  We first 
note that $\lambda_{1}^{\prime \alpha a}$ shifts under supersymmetry, while
$\lambda_{2}^{\prime \alpha a}$ and $\lambda_{i}^{\prime \alpha a}$ do not.
Therefore $\lambda_{1}^{\prime \alpha 1}$ and $\lambda_{1}^{\prime \alpha 2}$ 
are the would-be Goldstone fermions that arise from the supersymmetry breaking.  In unitary gauge, these fields vanish, and eqs.~(\ref{eqvirio2}) become
\bea
({\gamma^\mu})_{\alpha\beta} {\partial_\mu}\chi_\nu^{\prime\, \beta a}
+m\chi_\nu^{\prime\, \alpha a}&=&0 \nn\\[1mm]
{\partial^\mu}{\chi^{\prime\,\alpha a}_\mu}
-m({\gamma^\mu})^{\alpha\beta}{\chi^{\prime\, a}_{\beta\mu}} &=& 0 \nn\\[1mm]
(\gamma^\mu)_{\alpha\beta}  \partial_\mu
{\lambda}^{\prime\, \beta a}_{i}
-3m {\lambda^{\prime\, a}_{\alpha i}}&=&0 \nn\\[1mm]
({\gamma^\mu})_{\alpha\beta} {\partial_\mu}\lambda_{2}^{\prime\,\beta a}
-3m\lambda^{\prime\, a}_{2\alpha}&=&0.
\label{eqvirio3}
\eea
These are nothing but the equations of motion for two massive gravitini and six massive dilitini.  The two gravitini have eaten the two would-be Goldstone fermions, as required by the superHiggs effect.

\section{Conclusions}

In this paper we illustrated the superHiggs effect in heterotic string theory.  We first turned on a background NS-NS field $B_{\mu\nu}$, and deformed the BRST operator consistent with superconformal invariance.  We then derived the string-theory equations of motion for the background, as well as for the gravitino and the dilatino fields.  We found the spacetime supersymmetry generator and used it to derive the supersymmetry transformations of the spacetime fields.

We then studied a model in which spacetime is compactified on $M^7 \times T^3$, with a constant flux in the compact dimensions.  We showed that the non-zero $B_{\mu\nu}$-field spontaneously breaks supersymmetry.  We demonstrated that the would-be Goldstone fermions can be eliminated by a supersymmetry transformation, and that in the unitary gauge, the gravitini and the remaining six dilatini obey massive equations of motion.  In this way we illustrated the superHiggs effect in the full string theory, and not just in the effective field theory that arises at low energy.

\section*{Acknowledgments}

I.\ G.\ would like to thank J.\ Liu, B.\ Morariu 
and M.\ Porrati for useful discussions.
This work was supported in part by the National Science Foundation, grant NSF-PHY-0401513, and by the Department of Energy, contract
DE-FG02-91ER40651-TASKB.

\end{document}